# Non-Reciprocal and Collimated Surface Plasmons in Drift-biased Graphene Metasurfaces

D. Correas-Serrano and J. S. Gomez-Diaz*

Department of Electrical and Computer Engineering, University of California Davis, One Shields Avenue, Kemper Hall 2039, Davis, CA 95616, USA

*jsgomez@ucdavis.edu

*We explore the unusual non-reciprocal and diffraction-less properties of surface plasmon polaritons propagating in drift-biased graphene-based metasurfaces. We show that applying a drift-current on a graphene sheet leads to extremely asymmetric in-plane modal dispersions from terahertz to infrared frequencies, associated with plasmons with low-loss (high-loss and ultra-high confinement) traveling along (against) the bias. Strikingly, truly unidirectional wave propagation is prevented by the intrinsic nonlocal response of a graphene, a mechanism that shapes the energy flow over the surface. We also show that highly-directive hyperbolic plasmons completely immune to backscattering propagate obliquely along the drift in nanostructured graphene. Finally, we discuss how spin-orbit interactions can be exploited in this platform to efficiently launch collimated plasmons along a single direction while maintaining giant non-reciprocal responses. Our findings open a new paradigm to excite, collimate, steer, and process surface plasmons over a broad frequency band.*

The Lorentz reciprocity principle constrains the performance of photonic systems by enforcing identical responses when observation and excitation points are interchanged [1-4]. Recently, non-reciprocal surface plasmons-polaritons (SPPs) have merged the concept of one-wave optical propagation with the confinement and manipulation of light in engineered surfaces much smaller than the wavelength [5-10]. The possibility to collimate and dynamically control the direction of these waves within a surface opens exciting venues to realize miniaturized all-photonic integrated systems [11] and may enable new applications in sub-diffractive nanophotonics, including sensing, imaging, and computing. Even though non-reciprocal SPPs can be obtained taking advantage of magneto-optical effects [8, 12-20], the unavoidable need of magnetic materials under strong bias fields significantly lessens the practical interest of this solution. Alternative approaches able to break Lorentz reciprocity, usually relying on nonlinear phenomena [21-27], opto-mechanical interactions [28-31], or spatiotemporal modulations [32-38], could in principle be extended to create non-reciprocal devices for flatland photonics systems exploiting the enhanced light-matter interactions and extreme directionality that anisotropic and hyperbolic metasurfaces provide [39-47]. Unfortunately, combining the precise interference of a continuum of SPPs with different wavevectors required to collimate surface waves over engineered surfaces with the constrains imposed by non-reciprocal mechanisms is a challenging task. As a result, non-reciprocal SPPs have mostly been studied in simplified 2D waveguiding scenarios that neglect the potential excitation and propagation of surface waves along all directions within the surface.

A different approach to obtain non-reciprocal SPPs consists on applying drift current bias to the host surface [9, 48-54]. The effect of this drift current can be qualitatively understood as follows: since SPPs are collective charge oscillations coupled to light, they are strongly affected by these drifting charges and are either dragged or opposed by it, which causes guided waves to effectively see different media when propagating with or against the drift. Even though the drift velocities required to achieve strong non-reciprocal responses are difficult to obtain in most semiconductors and metals [48, 55], graphene has recently opened new possibilities in this context [49] thanks to its ultra-high electron mobility [56]. Indeed, drift velocities close to the Fermi velocity ($v_F \approx 10^8$ cm/s) have been experimentally reported in graphene samples suspended in free-space [57], applying rapid bias pulses [58], and in graphene encapsulated in hexagonal boron nitride (hBN) [59]. Merging drift-biased non-reciprocity with the large light-matter interactions, tunable responses, and rich variety of directional topologies enabled by graphene-based metasurfaces from terahertz to infrared frequencies [60-64] may open new venues to collimate, steer, and process plasmons immune to backscattering. Such broadband platform is especially remarkable due to its simplicity and tunability, as the drift-bias not only controls the non-reciprocity strength but can also modify the available states supported by the metasurface and in turn its electromagnetic response.

Consider the graphene sheet depicted in Fig. 1a, where a longitudinal voltage $V_{DC}$ induces a drifting of electrons along the sheet with velocity $\vec{v}_d = v_d \hat{y}$. Using self-consistent quantum mechanical methods and neglecting the possible dependence of the drift velocity on the electron energy, References [48,53, 65] recently showed that in the presence of this drift, graphene's conductivity becomes nonlocal and can be written as

$$\sigma_d(v_d, k_y) = \frac{\omega}{\omega - k_y v_d} \sigma_g(\omega - k_y v_d), \tag{1}$$

where $k_y$ denotes the wavevector component along the drift direction (see Fig. 1a) and $\sigma_g$ is graphene's conductivity without the presence of drift currents or magnetic bias. Intuitively, the drift bias introduces a Doppler shift to the conductivity of an amount equal to $k_y v_d$, plus a multiplicative factor of $\frac{\omega}{\omega - k_y v_d}$ that might lead to a negative Landau damping in which SPPs take kinetic energy from the drifting electrons and thus are amplified [54]. The fact that the supported SPPs show a non-reciprocal response immediately follows from Eq. (1), since reversing the propagation direction equates to changing the sign $k_y$ and therefore the sign of the Doppler shift, i.e., $\sigma_d(k_y, \omega) \neq \sigma_d(-k_y, \omega)$. To date, such plasmons have been studied assuming invariance in the transverse plane axis ($k_x = 0$) thus considering drift-biased graphene as a 2D waveguide problem [49, 66]. While this family of SPPs may dominate if the excitation is *x*-invariant and the structure infinite, graphene is usually studied experimentally using small sources such as nanotip scatterers or quantum dots [64, 67-69] that do not fit this description. Future experiments over drift-biased graphene are likely to use a similar approach. Furthermore, taking full advantage of graphene nanoplasmonics in real applications [60, 61] requires moving beyond the two-dimensional simplification and considering realistic three-dimensional scenarios. In a related context, recent experiments in the absence of drift-bias have shown that the intrinsic nonlocal response of graphene may significantly impact the features of the supported plasmons [70]. Such response arises because the finite Fermi velocity of electrons cannot follow the quick field variations of SPPs with high $|\vec{k}|$-values close to and above $\sim 300 k_0$ [71-73], with $k_0$ being the free-space wavenumber. In the presence of a longitudinal DC bias, nonlocality may play even a more critical role and conform the properties of ultra-confined SPPs traveling against the bias. A similar behavior has recently been reported in SPPs propagating at metal-dielectric interfaces [74], where nonlocal effects are the underlying mechanism that prevent the complete unidirectionality of these waves in the presence of a magnet.

In this Letter, we explore the non-reciprocal and collimation properties of SPPs propagating on drift-biased graphene-based metasurfaces. To this purpose, we characterize graphene combining the presence of the drift-bias (as shown in Eq. (1) [54]) with a rigorous conductivity model [72] that takes the intrinsic nonlocal response of graphene into account in the frequency band where intraband contributions dominate. Then, we derive the dispersion relation of the supported plasmons and develop a nonlocal and anisotropic Green's function approach to illustrate wave propagation on realistic three-dimensional configurations. Our theory, detailed in [66], is based on replacing graphene's surface conductivity by a drift- and wavenumber-dependent anisotropic conductivity tensor for every plane wave in the angular spectrum representation of the fields. In the case of homogeneous graphene, the drift bias leads to eigenstates that are extremely asymmetric with respect to the direction of the applied bias, corresponding to low-loss SPPs along the drift and to ultra-confined plasmons with moderate-loss traveling in the direction against it. Remarkably, nonlocal effects prevent truly unidirectional wave propagation and shape the energy flow of the supported plasmons over the surface. We also show that hyperbolic and diffraction-free SPPs *completely immune to backscattering* are supported in nanopatterned graphene metasurfaces that are longitudinally biased. Lastly, we exploit the photonic spin Hall effect in conjunction with this bias scheme to efficiently launch ultra-collimated SPPs along a single direction while keeping extreme unidirectional responses.

Fig. 1b shows the isofrequency contour (IFC) of the SPPs supported by a graphene sheet embedded in hBN (with relative permittivity 3.9, as in [49, 59]) for several drift currents $\vec{v}_d = v_d \hat{y}$. The group velocity of such waves, qualitatively illustrated in Fig. 1b using arrows, is always perpendicular to the IFC and can be determined by the gradient of the dispersion relation as $\vec{v}_g = \nabla_{\vec{k}_t} \omega(\vec{k}_t)$, with $\vec{k}_t$ being the in plane SPPs wavenumber. In the absence of any biasing, graphene is an isotropic material that possesses a circular IFC as all supported SPPs exhibit identical characteristics independently of their direction of propagation. As in all isotropic materials, the Poynting vector and the wavevector of the supported waves are aligned. For $v_d = 0.5 v_F$, the drift-bias breaks the symmetry of eigenstates with positive and negative $k_y$ leading to an effectively anisotropic two-dimensional medium in which the wave- and Poynting- vectors are no longer parallel. Plasmons in the bottom $\vec{k}$-quadrants (propagating towards negative $y$) are significantly more confined and lossy than the positive counterparts, with the highest asymmetry occurring when $k_x = 0$ and corresponding to the SPPs studied in 2D scenarios that assumes $x$-invariance [49, 66]. Increasing the drift bias further boost the asymmetry of the IFC. However, such IFCs always exhibit a closed shape independently of the drift value and thus SPPs are still supported in every direction, including along $-y$ ($k_x = 0$). The lack of truly unidirectional wave propagation appears due to the intrinsic nonlocal response of graphene as the finite velocity of electrons $v_F$ cannot follow the increasingly quick variations of the plasmons when $v_F |k_y| \leq |\omega - v_d k_y|$, a behavior consistent with the case of non-reciprocal plasmons on metal-dielectric interfaces biased with a magnetic field [74]. Strikingly, the confinement of such waves may be larger than the one of SPPs supported by pristine graphene in the absence of bias $\sim 300 k_0$ [72], which we attribute to the reference frame of an external observer taken in our numerical simulations. We stress that nonlocality is one of the key mechanisms that determines the properties of drift-biased graphene plasmonics. If non-local effects were not rigorously accounted for, the IFC would exhibit an open shape that would prevent wave propagation in a range of directions close to $-y$ and would significantly modify the energy flow of the supported SPPs.

Figs. 1c-d show the *z*-component of the electric field on the graphene sheet generated by a *z*-oriented dipole located 35 nm above for drift velocities $v_d = 0.5 v_F$ and $0.85 v_F$, respectively. Results show that *most energy travels towards the $-y$ half space*. This occurs because the emitter is so close to graphene

that high-$\vec{k}$ evanescent waves barely decay before reaching the surface, and once there, they couple to SPPs propagating *along directions oblique and against to the* $\vec{v}_d$ (larger $\vec{k}$ states in Fig. 1b). For sufficiently large drift bias (Fig. 1d), the waves radiated by the emitter cannot efficiently couple to SPPs going along $-y$ due to their extreme wavenumber and thus most power couple and travel towards oblique directions against the drift. If one is not interested in oblique beams and only requires wave propagation towards $+y$, moving the dipole further away from graphene is enough to filter the high-$\vec{k}$ components that would couple to those directions. This separation, $z_{dip}$, can be controlled with a moving nanotip or with a dielectric spacer [75-78]. Fig. 1e-f illustrate this scenario for $z_{dip} = 100$ nm. As expected, the higher-$\vec{k}$ components are filtered by free-space and very weakly excite SPPs in the oblique directions; SPPs now predominantly propagate towards the $+y$ half-space. We note that this beam is weakly collimated and exhibits low-loss [66], desirable properties for long distance SPP propagation.

The concept of drift-biased graphene can be expanded to hyperbolic metasurfaces comprised of an array of deeply subwavelength strips [41]. The structure under study is shown in Fig. 2a. Each drift-biased graphene strip has a conductivity $\bar{\bar{\sigma}}_b$ given by Eq. (2) and the whole metasurface can be modelled as an homogenous surface through an effective conductivity tensor $\bar{\bar{\sigma}}_{MTS}^{eff}$ that takes drift currents and inter-strip capacitive coupling into account, as detailed in [66]. Even though this approximate method neglects weak spatial dispersion effects due to granularity [73] and potential edge imperfections [56, 60] it allows to easily isolate and quantify drift-current-related effects. For the sake of consistency, we will employ the same operation frequency as in the previous examples and graphene of similar characteristics.

Consider now an array of strips with unit-cell period $L = 50$ nm and strip width $W = 25$ nm embedded in hBN, as described in Fig. 2. Fig. 2b shows the IFC of the SPPs supported in this hyperbolic metasurface for the same drift current values as in Fig. 1b, where again arrows point the direction of energy flow. The hyperbolic dispersion of the surface enforces that wave propagation towards the $+x$ ($-x$) half-space is associated to SPPs with negative (positive) $k_x$ [42, 66]. The effect that increasing $v_d$ has on the IFC is relatively similar to the case of homogeneous graphene, with eigenstates exhibiting smaller (larger) $k_y$ when pointing along (against) the current. The key difference in drift-biased hyperbolic metasurfaces is that the effective conductivity along $x$ is dominated by the capacitive coupling between strips, whereas the effective conductivity along $y$ is inductive and strongly dependent on $k_y$. Such nonlocal and anisotropic behavior tailors the eigenstates of the system. Specifically, in the upper $k_y$ space, increasing the drift current slightly modifies the inductive response of the graphene strips, which has the effect of smoothly flattening the top hyperbola and collimating SPPs towards the *y* axis. Quite differently, in the lower $k_y$ space, increasing $v_d$ leads to extreme inductive responses, drastically narrowing the hyperbola and thus the range of supported $k_x$ until no eigenstate with negative $k_y$ exists, completely forbidding SPPs towards the $y<0$ plane for $v_d=0.85v_F$. The supported hyperbolic SPPs are therefore *immune to backscattering* and thus may make an ideal platform for plasmonic isolators. Moreover, even if negative-$k_y$ states exist (e.g., with $v_d=0.5v_F$), they exhibit extremely high loss [66] and, contrary to pristine graphene, propagation close to the *x*-axis is forbidden by the intrinsic anisotropy of the metasurfaces, preventing potentially undesired energy loss toward oblique directions. Fig 2.c-d show the SPPs launched by a *z*-oriented dipole located 100 nm above such hyperbolic metasurface with $v_d=0.5v_F$. Positive-$k_y$ states carry collimated power towards $+y$, as predicted by the IFC, whereas negative-$k_y$ are barely excited and decay rapidly. Increasing $v_d$ to $0.85v_F$ (Fig. 2d) increases further the asymmetry, completely forbidding propagation against the drift.

To quantify the degree of non-reciprocity achievable by drift-biased graphene metasurfaces, we define isolation between two arbitrary points $\vec{r}_0'$ and $\vec{r}_0$ as the ratio $|E_v(\vec{r}_0,\vec{r}_0')/E_u(\vec{r}_0',\vec{r}_0)|^2$, where $\vec{v}$ and $\vec{u}$ are the polarization directions at $\vec{r}_0'$ and $\vec{r}_0$, respectively. Given the complexity of the structures under analysis, the vast parameter space, and the presence of multiple preferential propagation directions, performing a comprehensive evaluation of the isolating capabilities of this platform is an extensive task. For the sake of illustration, we will focus here on the non-reciprocity between a source with dipole moment $\vec{p}=\hat{z}$ (C·m) and the *z*-component of the electric field induced in an observation point that is aligned with the maximum of the plasmonic beam at a distance of $0.05\lambda_0$, where $\lambda_0$ is the free-space wavelength. Fig. 3a evaluates the isolation of the drift-biased graphene sheet shown in Fig. 1 using the positions $\vec{r}_0' = z_{dip}\hat{z}$ and $\vec{r}_0 = \vec{r}_0' + 0.05\lambda_0\hat{y}$ versus the drift velocity and $z_{dip}$, with positive (negative) isolation associated to propagation favored along (against) the drift current. As expected, SPPs propagation is commonly favored along $v_d$, with isolations as large as +70 dB for merely $0.05\lambda_0$ propagation length. Note that isolation (in dB) increases roughly linearly with this distance, and so does loss, since both are associated to exponential decay [65]. Isolation is found to monotonically increase with $v_d$, whereas there are optimal values of $z_{dip}$ that depend on $v_d$ and frequency [66]. The field intensity reaching $\vec{r}_0$ depends strongly on $z_{dip}$ due to free-space filtering of evanescent waves and weakly on $v_d$, as co-directional drift currents decrease SPP's decay rate [65]. For small $z_{dip}$, isolation generally increases with frequency, due to graphene's dispersive inductance ($\text{Im}[\sigma_g] \propto 1/f \rightarrow \frac{k_{spp}}{k_0} \propto f$) [66], entailing broadband nonreciprocal responses. As $z_{dip}$ increases, however, free-space filtering of evanescent components balances and eventually dominates this trend, leading to lower isolation. Fig. 3b quantifies the level of non-reciprocity in the drift-biased hyperbolic metasurface described in Fig. 2. Giant isolation is once again observed, with values larger than in homogeneous graphene. Drift-biased hyperbolic metasurfaces exhibit stronger protection to backscattering due to the absence of obliquely directed states, allowing to separately tailor the local density of states and non-reciprocity through the metasurface parameters and $v_d$, respectively. Moreover, the ability to steer the ultra-collimated hyperbolic beams by changing $v_d$ may in fact be an important advantage of drift-biased hyperbolic surfaces, as isolation remains large for a wide range of $v_d$ [66]. This may be used, for instance, to fine tune emitter-receptor alignment or to scan the metasurface plane, allowing to probe different points separated by deeply subwavelength distances.

Spin-orbit interactions [79-83] can be combined with the platform proposed here to enhance the excitation directionality of drift-biased SPPs. To this purpose, the spin angular momentum of an electric dipolar source can be chosen to match the transverse spin of a specific subset of surface waves supported by the metasurface [65]. For simplicity, we maximize here spin-orbit locking to SPPs propagating in one $x$-half-plane and minimizing it in the other. Fig. 4a shows the $z$-component of the electric field launched by a source with dipole moment $\vec{p} = \hat{x} + i\hat{z}\ (C \cdot m)$, i.e., circularly polarized in the $x$-$z$ plane and with a spin angular momentum $\vec{S} = -\hat{y}$ [66], onto the homogeneous graphene sheet of Fig. 1 with $v_d = 0.85 v_F$. Due to the spin-locking, the source excites non-reciprocal surface plasmons propagating towards the $+x$ semi-space that share similar $-S_y$ spin. Fig. 4b illustrates the response of the hyperbolic metasurface described in Fig. 2 when it is excited by a source with dipole moment $\vec{p} = -\hat{x} + i\hat{z}\ (C \cdot m)$ and an angular spin $\vec{S} = +\hat{y}$. The radiated fields couple to SPPs propagating in the $+x$ semi-space, which due to the hyperbolic dispersion of the surface possess negative $k_x$ wavenumbers (see Fig. 2b) and positive $S_y$ spin [66]. Results confirm that SPPs in the left half-plane are not excited at all and energy travels along a single collimated beam, as intended. This configuration exhibits exciting properties for manipulating and steering collimated SPPs over a surface, combining extreme directionality and reflection-immunity.

In conclusion, we have explored unidirectional and collimated plasmons in drift-biased homogeneous and patterned graphene. We have shown that applying drift currents to a graphene sheet leads to extremely asymmetric modal dispersions over a broad frequency range, associated to low-loss plasmons in the $\vec{k}$-space parallel to the bias and to very confined and lossy SPPs in the $\vec{k}$-space opposite to the drift. Nonlocal effects have been found to be an important mechanism that prevents truly unidirectional wave propagation and shapes the SPPs energy flow over the entire structure. We have also shown that the intrinsic non-reciprocal response of this platform enables giant isolation even for small source-observation distances and discussed the importance of excitation schemes to maximize it. Then, we have put forward magnet-less non-reciprocal hyperbolic metasurfaces by patterning graphene into strips and applying a drift-current bias to them. The structure supports *hyperbolic and broadband SPPs that are immune to back-scattering*. Lastly, we have applied fundamental concepts of spin-orbit photonics to further enhance the directionality of the excited plasmons, demonstrating that all power may be concentrated along a single collimated and non-reciprocal beam. The platforms proposed here may not only enable giant and broadband isolation in deeply subwavelength plasmonics systems but also lead to novel avenues for tunable SPP-collimation and routing over ultrathin surfaces.

This work was supported by the National Science Foundation with a CAREER Grant No. ECCS-1749177.

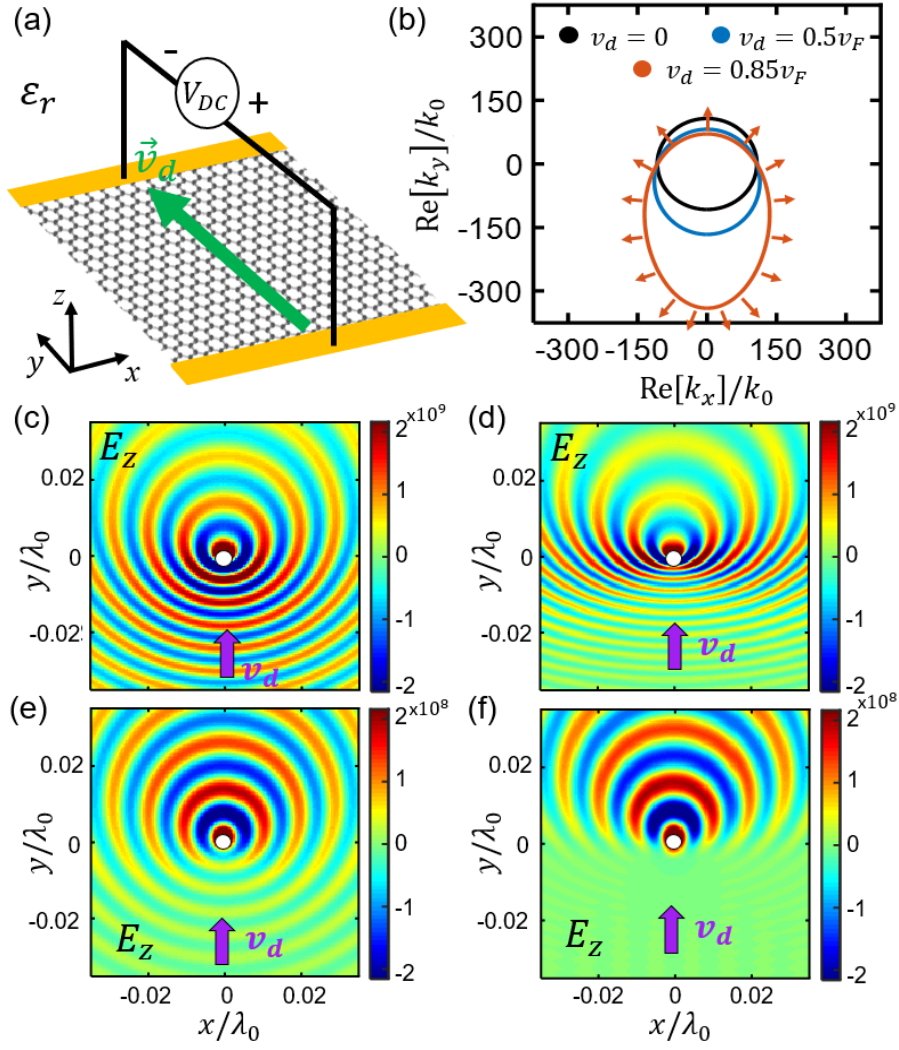

Fig. 1. Graphene sheet embedded in hBN and biased with a drift-current $\vec{v}_d = v_d \hat{y}$. (a) Schematic of the configuration. (b) Isofrequency contour of the structure for different drift currents $v_d$. Arrows shows the direction of energy flow (i.e., SPP group velocity). (c)-(f) $z$-component of the electric field (V/m) of the SPPs launched on the surface by a unitary point emitter with dipole moment $\vec{p} = \hat{z}$ (C · m) for various drift currents $v_d$ and dipole positions $\vec{r}' = z_{dip}\hat{z}$. (c) $v_d = 0.5v_F$ and $z_{dip} = 35$ nm. (d) $v_d = 0.85v_F$ and $z_{dip} = 35$ nm. (e) $v_d = 0.5v_F$ and $z_{dip} = 100$ nm. (f) $v_d = 0.85v_F$ and $z_{dip} = 100$ nm. Other parameters are $\mu_c = 0.2$ eV, $\tau = 0.5$ ps, T = 300 K, and a frequency of 21 THz.

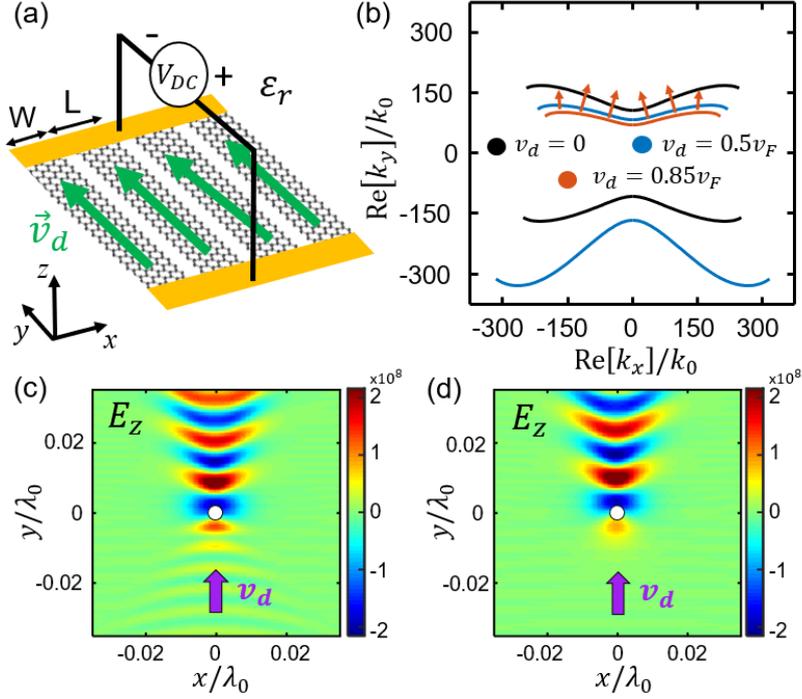

Fig. 2. Drift-biased hyperbolic metasurface comprised of tightly packed graphene strips with unit-cell period $L \ll \lambda_0$ [66] that is embedded in hBN. (a) Schematic of the configuration. (b) Isofrequency contour of the structure for different drift currents. Arrows shows the direction of energy flow (i.e., SPP group velocity). (c)-(d) z-component of the electric field (V/m) of the SPPs launched on the surface by a unitary point emitter with dipole moment $\vec{p} = \hat{z}$ (C·m) that is located at $\vec{r}' = z_{dip}\hat{z}$, with $z_{dip} = 100$ nm. Results are plotted versus the drift currents $\vec{v}_d = v_d\hat{y}$, with (c) $v_d = 0.5v_F$ and (d) $v_d = 0.85v_F$. The strip width and unit-cell period are set to $W = 25$ nm and $L = 50$ nm, respectively, with chemical potential $\mu_c = 0.4$ eV. Other parameters are as in Fig. 1.

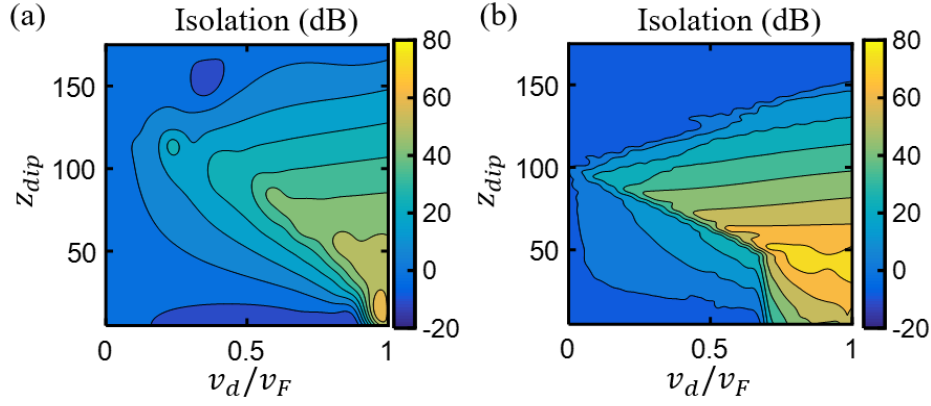

Fig. 3. Non-reciprocity in drift-biased graphene metasurfaces. (a) Isolation in the drift-biased graphene sheet described in Fig. 1 when $\vec{r}_0 = \vec{r}_0' + 0.05\lambda_0 \hat{y}$. (b). Isolation in the drift-biased hyperbolic metasurface described in Fig. 2 when $\vec{r}_0$ is chosen to be in the maximum of one hyperbolic beam while keeping $|\vec{r}_0 - \vec{r}_0'| = 0.05\lambda_0$ (see [66]). Results are plotted versus the velocity of the drifting electrons $v_d$ and the location of the $z$-polarized source/observation points over the metasurface, $z_{dip}$.

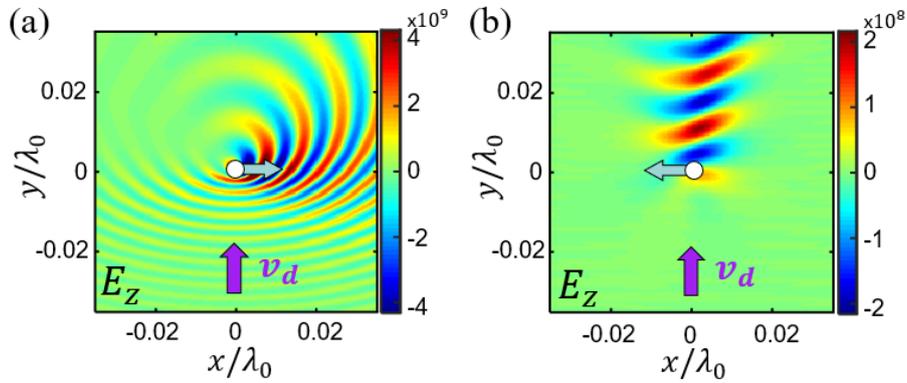

Fig. 4. $z$-component of the electric field (V/m) of SPPs launched by (a) an emitter with dipole moment $\vec{p} = \hat{x} + i\hat{z}$ (C · m) located at $z_{dip} = 35$ nm over the drift-biased graphene sheet described in Fig. 1; and (b) an emitter with dipole moment $\vec{p} = -2\hat{x} + i\hat{z}$ placed at $z_{dip} = 100$ nm above the drift-biased hyperbolic metasurface described in Fig. 2. The drift velocity is fixed to $v_d = 0.85 v_F$.